\documentclass[11pt]{article}

\usepackage[final]{acl}

\usepackage{amsmath}
\usepackage{graphicx}
\usepackage{xcolor}
\usepackage{amssymb}
\usepackage{listings}
\usepackage{booktabs}
\usepackage{amsmath}
\usepackage{amssymb}
\usepackage{mathtools}
\usepackage{amsthm}
\usepackage{bbm}
\usepackage{subfigure}
\usepackage{multirow}
\usepackage{hyperref}
\usepackage{subcaption}
\usepackage{natbib}
\usepackage{enumitem}

\theoremstyle{plain}

\theoremstyle{definition}

\theoremstyle{remark}

\DeclareMathOperator*{\concat}

\usepackage[T1]{fontenc}

\usepackage[utf8]{inputenc}

\usepackage{microtype}
\usepackage{url}
\usepackage{inconsolata}

\usepackage{graphicx}

\usepackage{soul}

\title{Program Structure-aware Language Models: \\ Targeted Software Testing beyond Textual Semantics}

\author{
 \textbf{Khang Tran},
 \textbf{Khoa Nguyen},
 \textbf{Cristian Borcea},
 \textbf{NhatHai Phan}
\\
 New Jersey Institute of Technology, Newark, NJ, USA
\\
 \small{{\{kt36, nk569, borcea, phan\}@njit.edu}
 }
}

\begin{document}
\maketitle

\begin{abstract}
Recent advances in large language models for test case generation have improved branch coverage via prompt-engineered mutations. However, they still lack principled mechanisms for steering models toward specific high-risk execution branches, limiting their effectiveness for discovering subtle bugs and security vulnerabilities. We propose \texttt{GLMTest}, the first program structure-aware LLM framework for targeted test case generation that seamlessly integrates code property graphs and code semantics using a graph neural network and a language model to condition test case generation on execution branches. This structured conditioning enables controllable and branch-targeted test case generation, thereby potentially enhancing bug and security risk discovery. Experiments on real-world projects show that \texttt{GLMTest} built on a Qwen2.5-Coder-7B-Instruct model improves branch accuracy from 27.4\% to 50.2\% on TestGenEval benchmark compared with state-of-the-art LLMs, i.e., Claude-Sonnet-4.5 and GPT-4o-mini.
\end{abstract}

\section{Introduction}

Testing is a cornerstone of modern software development, serving to validate program correctness and uncover functional defects before deployment \cite{battina2019artificial, wang2024software, wang-etal-2025-testeval}. It is equally critical for software security, as systematically generated test cases can reveal crashes, anomalous behaviors, and exploitable vulnerabilities \cite{liang2018fuzzing, zhu2022fuzzing}. Consequently, testing accounts for a substantial portion of software engineering effort, as reflected in industry reports \cite{KPMG2024TestingMarket}. The rising cost and complexity of software systems have therefore heightened the demand for automated test generation techniques that improve testing efficiency, effectiveness, and coverage \cite{brunetto2021introducing, baqar2025future}.



\textbf{Motivation.} Recent advances in large language models (LLMs) have enabled new approaches to automated test-case generation~\cite{harman2025mutation, 10172800, pan2025aster}. In particular, LLMs have been used to mutate test cases to expand execution coverage~\cite{harman2025mutation, 10172800, pan2025aster}. However, most existing methods rely on prompt-engineering heuristics, making the mutation process difficult to control due to LLM stochasticity and lacking principled optimization to target specific execution branches or high-risk code regions. Consequently, prompt-engineered test cases often fail to exercise security-critical paths, limiting their effectiveness in bug discovery~\cite{weissberg2024sok}. This highlights the need for explicitly optimized test-generation techniques that target high-risk execution branches.

\textbf{Challenges.} Developing LLM-based mechanisms for generating test cases targeting specific execution branches is challenging. Even with greedy sampling, the inherent stochasticity of LLMs \cite{astekin2024exploratory, song2025good} makes outputs difficult to control, often failing to reach the intended branches \cite{huang2025challenges, feng2025fuzzing}. Moreover, purely textual representations do not adequately capture dependencies among code objects, limiting the model’s understanding of program structure and execution behavior and hindering precise execution guidance.




\textbf{Our Solution.} We propose \textbf{\texttt{GLMTest}} to encode the program (under test) by transforming its graph representation and developer-provided textual information into a shared high-dimensional embedding space using a heterogeneous graph neural network (GNN) and an LLM. Unlike prior works~\cite{chen2025bridging, liu2025vul} that typically feed graph features into an encoder and then query an LLM only at the sequence level, \texttt{GLMTest} jointly trains a heterogeneous GNN and an LLM to learn node-level embeddings that are directly aligned with branch-specific execution masks and injected into the LLM as branch-conditioned inputs, explicitly tailoring the graph representation to targeted test case generation rather than generic code understanding. Thus, \texttt{GLMTest} provides a controllable mechanism for generating test cases that execute targeted program locations.



At inference time, \texttt{GLMTest} can generate test cases oriented toward specific targeted locations in the code, providing a practical way to exercise high-risk branches and potentially expose underlying defects or security risks. Furthermore, \texttt{GLMTest} can be applied in a coverage-oriented setting to systematically expand coverage for regression and coverage-driven testing pipelines. In both cases, \texttt{GLMTest} offers finer-grained control over which execution paths are exercised than prior prompt-engineered LLM approaches, enabling more precise and interpretable test case generation.


\textbf{Contributions.} Our contributions are as follows: \textbf{(1)} We present \texttt{GLMTest}, the first graph-enhanced language modeling framework for branch-targeted test case generation. By jointly modeling program structure and textual semantics, \texttt{GLMTest} enables focused testing of high-risk branches and systematic exploration to improve branch coverage. \textbf{(2)} We also introduce a new dataset derived from real-world repositories and a training strategy that learns fine-grained, branch-oriented embeddings for targeted test generation\footnote{Our implementation and dataset can be found here: \url{https://github.com/khangtran2020/glmtest}}. \textbf{(3)} Experiments on Python programs from the TestGenEval benchmark show that \texttt{GLMTest} significantly outperforms enterprise LLMs (e.g., Claude-Sonnet-4.5), improving branch accuracy from 27.4\% to 50.2\% while achieving high branch coverage.






\section{Background \& Related Work}




\textbf{LLMs for Test Case Generation}. LLMs have become central to automated code generation, improving software development workflows~\cite{parvez-etal-2018-building}. Trained on large-scale open-source code and fine-tuned for instruction following~\cite{roziere2023code}, they have recently been applied to software testing to generate readable, correct test suites with improved coverage~\cite{tufano2021unittestcasegeneration}. Existing approaches fall into two categories: fine-tuning and prompt engineering. Fine-tuning methods specialize LLMs for test generation using curated code--test pairs~\cite{tufano2021unittestcasegeneration, ALAGARSAMY2024107565, 10.1109/ASE56229.2023.00193}, while prompt-based methods guide frozen LLMs with structured program features (e.g., signatures and control flow) to achieve coverage-oriented test generation~\cite{, 10.1145/3661167.3661216, 10.1145/3663529.3663801, DAKHEL2024107468}.

\textbf{Code Property Graph (CPG).} In software analysis, graph representations encode relationships among program elements and serve as structured inputs for program analysis~\cite{10.1145/3664649}. Common examples include abstract syntax trees for syntactic structure~\cite{7582748}, control-flow graphs for execution paths~\cite{10.1145/3236024.3236068}, and data-flow graphs for data dependencies~\cite{10.5555/1965094}. Recent work~\cite{lekssays2025llmxcpg, chen2025bridging} explores combining code graphs with LLMs, mainly using graphs as auxiliary knowledge to enrich prompts rather than tightly integrating program structure and code semantics into an optimized model for test generation~\cite{10.1145/3643769}. More details can be found in Appendix~\ref{appx:related-work}.

\section{Problem Formulation}

\paragraph{CPG Annotation.} The CPG can be defined as $G = (V, E)$, where $V$ is the set of nodes and $E$ is the set of edges. Each node $v \in V$ corresponds to a program element (e.g., a statement, expression, variable, or function) and is associated with a feature vector $x_v \in \mathbb{R}^d$. This vector can encode multiple static attributes, such as the tokenized code snippet, syntactic type (e.g., assignment, call, branch), and program location (e.g., file, line, and column). By stacking all node features, we obtain the node feature matrix $X \in \mathbb{R}^{|V| \times d}$. The edge set $E$ is partitioned into subsets $E = \bigcup_j^r E_j$, where $E_j$ contains edges of type $j$ (e.g., abstract syntax tree) and $r$ is the total number of edge relations. In this way, the CPG represents the program as a heterogeneous, multi-relational graph that captures the program's structural information.

\begin{figure}[t]
\centering\includegraphics[width=0.8\linewidth]{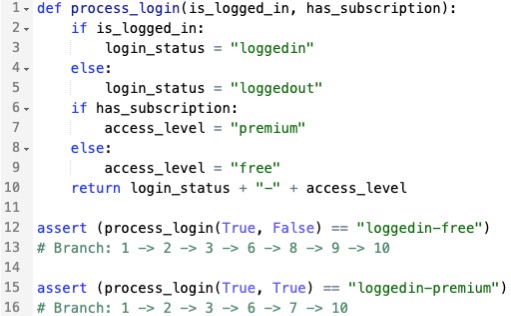}
    \caption{A Python function example annotated with line numbers and branch paths. Two test cases (Lines 12 and 15) are shown with their corresponding execution branches (Lines 13 and 16), illustrating how different input combinations traverse distinct branches.} \vspace{-10pt}
    \label{fig:branch-example}
\end{figure}

\textbf{Setting of \texttt{GLMTest}.} Given a program $S$, the test case generator produces a test suite $\tau=\{t_i\}_{i \in [1,n]}$, where $n$ is the number of generated test cases. An \emph{execution branch} is the control-flow path the program takes for a given input, i.e., the ordered sequence of executed decisions and statements executed for a test case \cite{ammann2008introduction}. Thus, we consider that by executing test case $t_i\in\tau$ on $S$ we can extract an execution branch $\hat{b}_i$ as the ordered sequence of statements executed by $S$ under test case $t_i$. Figure~\ref{fig:branch-example} illustrates two test cases for \texttt{process\_login}: the first call in line 12 drives execution along the branch indicated in line 13, while the second call in line 15 follows the branch indicated in line 16. We denote $\hat{b}_i = \texttt{Exec}_S(t_i)$ as the process of executing $t_i$ on $S$ which returns the execution branch $\hat{b}_i$. We note that two test cases $t_i, t_j \in \tau$ can derive the same execution branch, i.e., $\hat{b}_i = \hat{b}_j$ since they can produce the same control-path of the program. Let $B_S=\{b_i\}_{i \in [1, m]}$ denote the set of all possible execution branches of $S$, where $m$ is the total number of branches. The conventional coverage-driven objective of test case generation is to generate a test suite whose induced branches maximize coverage over possible execution branches $ B_S$.

\textbf{Goals.} In this work, we consider an objective tailored to branch-targeted test case generation while remaining aligned with the conventional coverage-driven objective. Specifically, \texttt{GLMTest} trains a model $f_{\theta}$ that takes as input the program $S$ and a target execution branch $b$ indicated by the developers, and outputs a test case $\hat{t}$. The training objective is to maximize the probability that executing the generated test on $S$ realizes the target branch:

\begin{equation}
\small
     \theta^* = \arg\max_{\theta}\sum_{b \in B_S}\Pr\Big[\texttt{Exec}_s(\hat{t}) = b | \hat{t} = f_{\theta}(S, b)\Big]. \label{eq:goal}
\end{equation}

By optimizing this objective, \texttt{GLMTest} is tailored to generate test cases whose execution on $S$ will align with the targeted execution branch. Furthermore, this also allows \texttt{GLMTest} to iterate over branches in $B_S$ and synthesize a test suite that improves coverage over $B_S$, thereby maximizing the execution branch coverage.

\section{\texttt{GLMTest} Framework}

\begin{figure*}[t]
\centering\includegraphics[width=\linewidth]{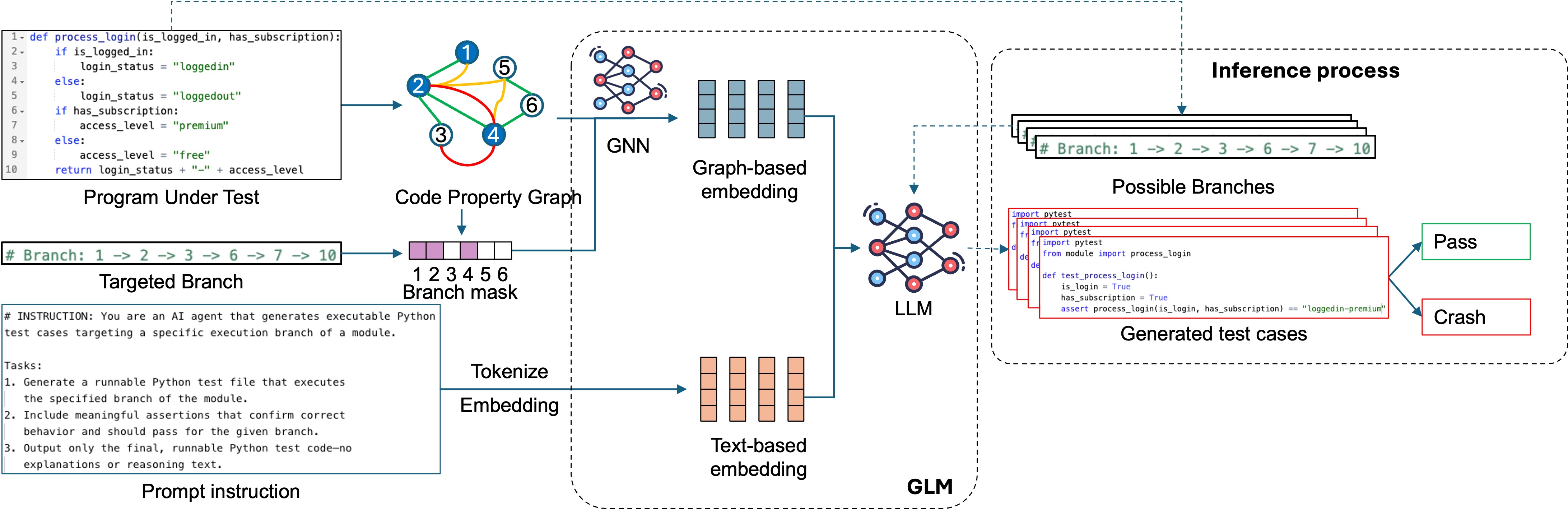}
    \caption{\texttt{GLMTest} pipeline.}
    \label{fig:pipeline}
\end{figure*}


This section describes \texttt{GLMTest} in detail with its training and inference processes.

\subsection{Overview}
 
Figure \ref{fig:pipeline} illustrates the operation pipeline of \texttt{GLMTest} framework, which seamlessly integrates code structural information with code textual information using a GNN and an LLM, as follows. \textbf{(1)} The GNN extracts the code structural information of the targeted execution branch. Unlike prior graph–LLMs, our GNN module is optimized along with the LLM to induce structural embeddings aligned with targeted branches. \textbf{(2)} The LLM combines the structural embeddings with the text embedding of the instruction to generate test cases executing targeted branches. By incorporating both structural and textual information, the LLM learns branch-oriented representations more effectively, steering the model generation toward relevant execution paths.

First, the framework extracts a CPG $G$ that captures the control-flow and data-dependency relationships of program $S$. Then, for a targeted execution branch $b$, it induces a specific set of nodes $V_b \subseteq V$ in $G$ that are components related to execution statements in $b$ (based on their location in $S$). We represent this set as a branch mask $m \in \{0, 1\}^{|V|}$, where $m_i = 1$ if $i \in V_b$; otherwise, $m_i = 0$. Then, the GNN module $f_g$ takes the CPG $G$ and the branch mask $m$ as input, and derives branch-aware structural embeddings that capture the structural dependencies of $b$.

Second, \texttt{GLMTest} adopts a text-based instruction prompt that specifies the testing objective and format, providing the LLM with high-level guidance on constructing effective test cases to exercise the targeted branch. The prompt is tokenized and encoded into textual embeddings. Then, the structural embeddings are concatenated with the textual embeddings, which are passed to the LLM module $f_{lm}$ to generate executable test cases executing $b$.


We train \texttt{GLMTest} end-to-end on a high-quality dataset curated from real-world projects with human-written test cases. For program $S$, we extract its human-written test cases and their associated execution branches. Then, $f_{lm}$ and $f_g$ are jointly trained with a supervised learning method to generate test cases from the instruction prompt text embedding and the branch-aware structural embedding. \texttt{GLMTest}'s flexible structure allows both $f_{lm}$ and $f_{g}$ to simultaneously train under conventional gradient-based optimizers, optimizing the same objective for targeted test case generation.

At inference time, developers can specify execution branches to test, and \texttt{GLMTest} generates test cases explicitly aimed at exercising those branches. In practice, developers can specify the execution locations (e.g., code lines, code blocks, or functions) from which GLMTest will automatically extract and derive execution branches. In addition, \texttt{GLMTest} can adapt to coverage-driven settings by enumerating feasible branches that are detectable by static analyzers and generating test cases across them. The resulting test cases are executed on the program, and their outcomes are used assess the program's execution~\cite{song2019sok}.

\subsection{Model Structure of \texttt{GLMTest}}





Let us describe the graph language–modeling module, the core component of \texttt{GLMTest}, to integrate structural and textual information from the program $S$ for test case generation.

\textbf{Execution Branch Embeddings}. We first introduce the GNN module of \texttt{GLMTest}, which extracts branch-aware structural embeddings from the CPG $G$. We employ a $K$-layer heterogeneous GNN \cite{schlichtkrull2018modeling} to capture different types of relations among nodes in $G$. Each layer $k \in [1, K]$ takes the node embeddings $h_v^{k-1}$ for $v \in V$ from the previous layer $k - 1$ and updates them as follows:
\begin{align}
    h^{k}_{N_j(v)} &= \operatorname{AGG}\big(h^{k-1}_v \cup \{h_u^{k-1} : u \in N_j(v)\}\big), \nonumber \\ 
    h^{k}_{v,j} &= \sigma\big(h^{k}_{N_j(v)}, W^{k}_j\big), \nonumber
\end{align}
where $N_j(v)$ is a set of neighborhood nodes of $v$ under relation type $j$ with edge set $E_j$, $h^{0}_v = x_v$ is the initial node feature, $\operatorname{AGG}(\cdot)$ is an aggregation function, $W^{k}_j$ is the trainable parameter matrix of layer $k$ for relation $j$, and $\sigma(\cdot)$ is a message-passing function (e.g., graph attention~\cite{velivckovic2017graph} or GraphSAGE~\cite{hamilton2017inductive}). 

This heterogeneous GNN propagates information along different relations in the CPG so that each node embedding captures its local program structural information. Also, the GNN backbone is modular, allowing \texttt{GLMTest} to benefit from future advanced GNN structures.

At the last layer $K$, the GNN component aggregates relation-specific embeddings into an overall embedding using a pooling function $\texttt{pool}(\cdot)$, e.g., a summation or average pooling operator, as follows: $h_v^K = \texttt{pool}\big(\{h_{v,j}^K\}_{j=1}^r\big).$
This step yields a unified embedding $h_v^K \in \mathbb{R}^{d_h}$ integrating information propagated through all edge relations $r$, where $d_h$ is the hidden dimension. Then, the embeddings of the targeted branch $b$ are derived by stacking the set of node embeddings related to the targeted branch $b$: $e_b = \texttt{stack}\Big(\{h_v^K\}_{v\in V_b}\Big) \in \mathbb{R}^{|V_b|\times d_h}$. This set of branch embeddings encodes the structural context of all nodes participating in the targeted branch, containing fine-grained information along the execution path.

\textbf{LLM Module.} \texttt{GLMTest} adopts a prompt template tailored to the TestGenEval benchmark~\cite{jaintestgeneval} (Figure~\ref{fig:prompt-template}, Appendix~\ref{appx:supplemental_results}) explicitly defining the model’s role and the prompt's inputs, and it constrains the output to a runnable and valid test case. The prompt's inputs include: \textbf{(i)} the program's source code, \textbf{(ii)} the execution-branch information (line executed), \textbf{(iii)} the importable program's path, \textbf{(iv)} the branch embeddings, and \textbf{(v)} a code snippet showing how to import the program. 

To combine the structural embeddings with textual embeddings, we introduce a \textit{graph token} \texttt{\textless|graph\_pad|\textgreater}, which is included in place of item (iv) as a placeholder for the branch embeddings. The branch embeddings $e_b$ are then integrated by replacing the embeddings of the graph tokens with $e_b$, yielding an input embeddings $e_{inp}$, which is forwarded through the LLM $f_{lm}$ to produce the token-level logits for next-token prediction. Thus, the structural signal influences all subsequent decoding steps, guiding the model toward generating test cases that exercise the targeted branch. 

\subsection{Training}
\label{sec:train}

\paragraph{Data Curation.} Since no existing dataset is tailored to branch-targeted test case generation, we curate supervision signals directly from real-world projects and their developer-written test suites. For each program $S$, we first collect its existing test suite $\tau$ written by developers and decompose it into individual test cases $\{t_i\}_{i=1}^{|\tau|}$. We process each test case $t_i$ to achieve high-quality requirements by removing unnecessary imports and dead code from each test set, ensuring $t_i$ focuses only on the targeted branch and reducing hallucination. 

We then execute each test case $t_i$ and record the corresponding branch $\hat{b}_i = \texttt{Exec}_s(t_i)$ executed in $S$. This branch information is used to build the input prompt and CPG-based structural features, while the original test case serves as the ground-truth target output. This automatic procedure yields realistic (program, branch, test case) triples aligned with the training objective in Eq. \eqref{eq:goal}. Training on this dataset guides the model to generate accurate test cases for the targeted execution branch, resulting in executable, branch-aware test cases. 

For each executed test case, we construct a training sample by extracting the CPG $G$ of $S$, deriving the branch mask $m_i$ associated with $b_i$, and instantiating the corresponding instruction prompt $p_i$, yielding a dataset
$D = \{(G, p_i, m_i, t_i)\}_{i=1}^{|D|}$. We release this branch-annotated dataset to support and encourage future research on structure-aware, branch-targeted test case generation.

\paragraph{Training Objectives.} The \texttt{GLMTest} model is then trained to optimize the following objective:
\begin{align}
\small
    \hat{t}_i &= f_{\mathrm{lm}}\big(e_{p_i}, f_g(G, m_i)\big), \\
    \theta^* &= \arg\min_\theta \sum_{i = 1}^{|D|} \ell(t_i, \hat{t}_i) + \lambda\|\theta\|_2,
\end{align}
where $\theta$ denotes all model parameters (including those of $f_g$ and $f_{\mathrm{lm}}$), $\lambda$ is an $\ell_2$ regularization coefficient, and $\ell(\cdot, \cdot)$ is the token-level training loss (e.g., cross-entropy). 

Because the branch embedding is concatenated into the embedding $e_{p_i}$, this operation remains fully differentiable, allowing gradients to backpropagate through the GNN $f_g$. As a result, the entire \texttt{GLMTest} pipeline can be trained end-to-end via gradient-based optimization (e.g., Adam). In practice, this pipeline can be instantiated under different training paradigms, such as supervised fine-tuning (SFT) on developer-written test cases or reinforcement learning from human feedback (RLHF) \cite{patil2024review} to further align generations with preferred testing behaviors. In our experiments, we focus on SFT due to its stability for large-scale training and leave RLHF-based refinement for future work.

 


\section{Experimental Results}

We conduct an extensive experiment to evaluate \texttt{GLMTest} around three research questions:

\textbf{RQ1:} Can \texttt{GLMTest} effectively generate test cases that exercise a targeted branch compared to state-of-the-art LLMs?

\textbf{RQ2:} Does \texttt{GLMTest} generate high-quality test suites that achieve competitive coverage?
    
\textbf{RQ3:} What is the contribution of \texttt{GLMTest}'s components to its overall performance?


\subsection{Experiment Setup}

\textbf{Datasets.}
We base our experiments on the TestGenEval dataset~\cite{jaintestgeneval}, a large-scale dataset for evaluating unit test case generation and completion. TestGenEval is constructed from SWEBench and comprises 68{,}647 test cases paired with 1{,}210 modules with executable Docker environments. In our setting, we treat individual Python modules within a repository as programs under test, and the associated developer-written test cases as ground-truth test cases for these programs. We decompose each test cases and spurious dependencies. We then execute each test case and collect branch information. Each test case and its associated set of executed branches form a data point, yielding a triplet \{program, branch, test case\} as described in Section~\ref{sec:train}. Our processing results in 40,868 data points illustrated in Table~\ref{tab:dataset-stats} (Appendix \ref{appx:supplemental_results}). We reserve 1{,}344 instances for evaluation, sampled uniformly across projects, ensuring that programs in the test set do not appear in the training set. Full details of our dataset are in Appendix \ref{appx:dataset}.


\textbf{Implementation.} To construct a CPG, we use Joern~\cite{yamaguchi2014modeling} for each program, and represent each node with a 772-dimensional feature vector obtained by concatenating a 768-dimensional \texttt{codet5p-110m-embedding} code embedding with a 4-dimensional encoding of categorical attributes (node type, order, and location). For branch masks, we build a binary branch mask by aligning line ranges with nodes. If no node aligns with a branch, we fall back to a special \emph{``not available''} structural input. This design provides an interpretable mapping from dynamic execution to static structure while remaining compatible with standard coverage tools. We employ \texttt{GLMTest} instantiated with a 3-layer graph attention network (GAT) using 8 attention heads per layer and Qwen2.5-Coder-7B-Instruct~\cite{hui2024qwen2} as the backbone LLM. Full details are in Appendix~\ref{appx:implementation}.

\textbf{Metrics.} We evaluate \texttt{GLMTest} using three complementary metrics: (i) \textbf{\texttt{Pass@1}}~\cite{jaintestgeneval} reflects the basic \emph{functional correctness} and executability of the generated test cases; (ii) \textbf{Branch Coverage} (\texttt{BranchCov})~\cite{wang2024software} measures how many feasible execution branches we can explore, and thus reflects the \emph{``testing utility''} of the generated suite; (iii) \textbf{Branch Accuracy} (\texttt{BranchAcc}) measures the success in executing the targeted execution branches; and (iv) \textbf{Branch Overlap} (\texttt{BranchOverlap}) measures the percentage of targeted branches that are covered by the generated test cases. It is worth noting that we modified the TestGenEval~\cite{jaintestgeneval} pipeline so that branch coverage is computed even if a generated test case fails due to incorrect assertions, and any executed branch is still recorded for coverage statistics.

As in TestGenEval~\cite{jaintestgeneval}, \texttt{Pass@1} measures the percentage of test cases in the generated test suite that pass when run on the program. \texttt{BranchCov} measures how much branch coverage we obtain from these generated test cases. Specifically, for each program, we run the subset of generated test cases whose execution and assertions succeed, compute branch coverage, and report the average fraction of covered branches across the programs.

\texttt{BranchAcc} captures whether the ground-truth targeted branch $b_i$ is exercised by the generated test (with executed branch set $\hat{b}_i$), and \texttt{BranchOverlap} measures how much of $b_i$ is actually covered, defined on a dataset $D$:
\begin{align}
    &\texttt{BranchAcc} = \frac{1}{|D|}\sum_{i=1}^{|D|} \mathbb{I}[b_i \in \hat{b}_i], \nonumber \\
    &\texttt{BranchOverlap} = \frac{1}{|D|}\sum_{i=1}^{|D|}\frac{|b_i \cap \hat{b}_i|}{|b_i|}, \nonumber
\end{align}
where $|b_i|$ is the number of statements in $b_i$ and $|D|$ is the size of the dataset $D$.


\textbf{Baselines.} There is a growing body of work on LLM-based test case generation \cite{harman2025mutation, ryan2024code}, among which only ASTER~\cite{pan2025aster} and CodaMOSA~\cite{10172800} directly target Python test case generation. However, both mechanisms are tightly coupled to Pynguin and project-specific import configurations, requiring modules to be directly importable from the local filesystem. This setting is incompatible with the containerized, repository-level setup of the TestGenEval dataset, where programs are executed in pre-built Docker environments. Our efforts to adapt and reproduce these methods in TestGenEval ultimately proved inapplicable to TestGenEval's containerized setting. Therefore, we focus on two strong and controllable LLM-based baselines that are fully compatible with TestGenEval. (i) \textbf{Prompt Engineering} (PE): we follow the TestGenEval protocol and prompt templates to query state-of-the-art commercial LLMs (Claude-Sonnet-4.5 and GPT-4o-mini). (ii) \textbf{Fine-tuning} (FT): we fine-tune the same backbone LLM used in \texttt{GLMTest} on the \{program, targeted branch, test case\} triples, but remove the GNN component and feed only textual instruction. We use the same LLM as \texttt{GLMTest} without the GNN component and mark all branch embeddings as \emph{Not Available}, so the model receives only textual inputs. More details are in Appendix \ref{appx:baseline-details}

\subsection{RQ1: Can \texttt{GLMTest} effectively generate test cases that exercise a targeted branch compared to state-of-the-art LLMs?}


\begin{figure}[t]
    \centering
    \includegraphics[width=\columnwidth]{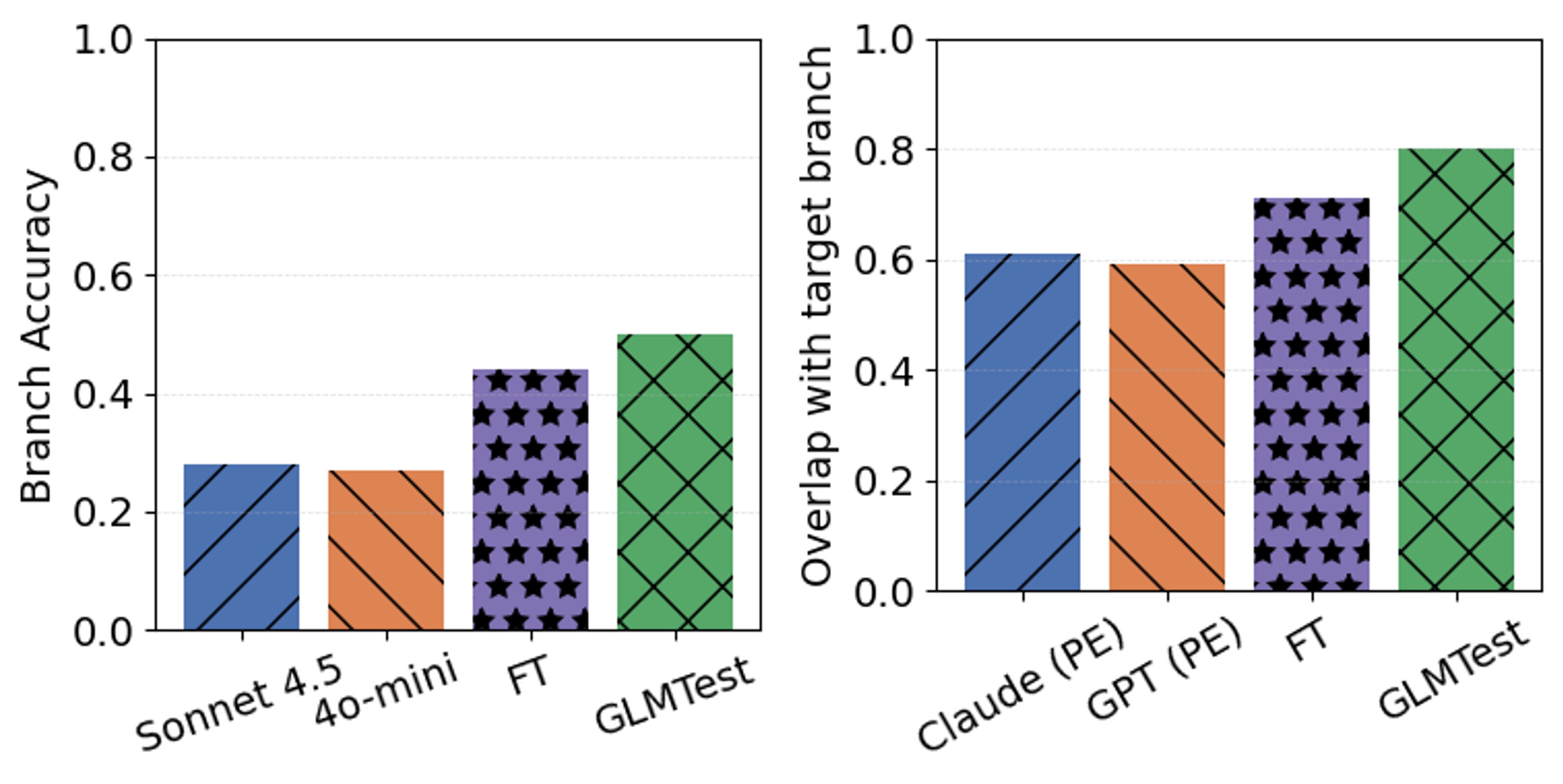}
    \caption{Branch accuracy and branch overlap with the targeted branches of \texttt{GLMTest} and baselines.}
    \label{fig:branch-metrics}
\end{figure}


Compared to the prompt-engineering (PE) baselines built on Claude-Sonnet-4.5 and GPT-4o-mini, \texttt{GLMTest} substantially improves the ability to exercise the targeted branches (Figure~\ref{fig:branch-metrics}): overall branch accuracy increases from 0.274 (GPT-4o-mini) and 0.292 (Claude-Sonnet-4.5) to 0.502 (\texttt{GLMTest}), registering a relative performance improvement of 71.9\%. It is worth noting that \texttt{GLMTest} uses a small, open-source model (Qwen2.5-Coder-7B-Instruct) augmented with our branch-structured conditioning rather than relying on massively scaled LLMs such as Claude-Sonnet-4.5 and GPT-4o-mini. Similar results are observed for \texttt{BranchOverlap}. Specifically, \texttt{BranchOverlap} increases on average from $0.615$ of the PE baselines to $0.794$ of \texttt{GLMTest}, indicating that \texttt{GLMTest} is more likely and more consistent to reach the targeted branch across tasks. Per-repository results mirror this overall performance. For instance, on \texttt{django} repository, compared with test cases generated by Claude-Sonnet-4.5, branch accuracy improves from $0.46$ to $0.68$ with \texttt{GLMTest}, and the fraction of targeted branches covered by the generated test cases increases from $0.63$ to $0.91$.

Finally, comparing \texttt{GLMTest} with FT isolates the effect of the GNN component. \texttt{GLMTest} improves \texttt{BranchAcc} from $0.44$ (FT) to $0.50$ and \texttt{BranchOverlap} from $0.71$ to $0.80$. Similar patterns appear in complex repositories, such as \texttt{xarray}, where \texttt{GLMTest} improves \texttt{BranchAcc} from $0.00$ (FT) to $0.61$ and \texttt{BranchOverlap} from $0.15$ to $0.85$, suggesting that code graph-based structural embeddings provide a rich signal that helps the model localize and execute the targeted execution branches.



\begin{figure}[t]
    \centering
    \includegraphics[width=\columnwidth]{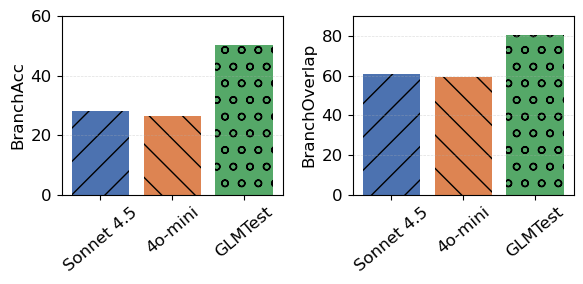}
    \caption{Branch accuracy and branch overlap with the targeted branches of \texttt{GLMTest} vs. baselines with execution feedback.}
    \label{fig:branch-metrics-fb}
\end{figure}

\begin{figure}[t]
    \centering
    \includegraphics[width=\columnwidth]{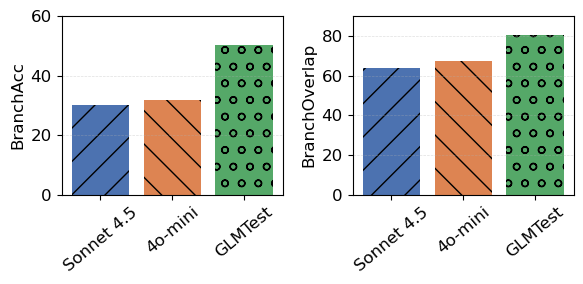}
    \caption{Branch accuracy and branch overlap with the targeted branches of \texttt{GLMTest} vs. RAG augmentation baselines.} \vspace{-5pt}
    \label{fig:branch-metrics-rag}
\end{figure}

\textbf{Advanced Prompting Techniques}. We compare \texttt{GLMTest} with an execution feedback baseline under the same branch-targeted evaluation protocol (Figure \ref{fig:branch-metrics-fb}). Specifically, we provide the generated test case and its associated execution branch, and ask the models to revise their outputs accordingly. Under this setting, 4o-mini and Sonnet-4.5 achieve 26.5\% and 28.5\% \texttt{BranchAcc} with 59.5\% and 60.2\% \texttt{BranchOverlap}, respectively, which are lower than \texttt{GLMTest}'s 50.2\% \texttt{BranchAcc} and 80.2\% \texttt{BranchOverlap}. These results indicate that iterative execution feedback, while providing some guidance, remains significantly less effective than explicit structural conditioning at reliably satisfying branch-specific execution constraints, highlighting \texttt{GLMTest}'s advantage.

In addition, we compare \texttt{GLMTest} with a retrieval-augmented generation (RAG) baseline (Figure \ref{fig:branch-metrics-rag}), constructing the retrieval corpus from the \texttt{GLMTest} training data. Specifically, for each training instance, we encode the prompt using OpenAI's \texttt{text-embedding-3-small}. At inference time, the RAG baseline retrieves the top-3 most similar samples via cosine similarity and provides them, along with their associated human-written test cases, as in-context examples. The baseline is then prompted to generate test cases targeting the specified branches. Under this setting, 4o-mini and Sonnet-4.5 achieve 31.8\% and 30.1\% \texttt{BranchAcc} with 67.5\% and 64.0\% \texttt{BranchOverlap}, respectively. These results further confirm that in-context retrieval augmentation remains insufficient for reliably satisfying branch-specific execution constraints compared to the explicit structural conditioning employed by \texttt{GLMTest}.

\subsection{RQ2: Does \texttt{GLMTest} generate high-quality test suites that achieve competitive coverage?}

\begin{figure}[t]
    \centering
    \includegraphics[width=\columnwidth]{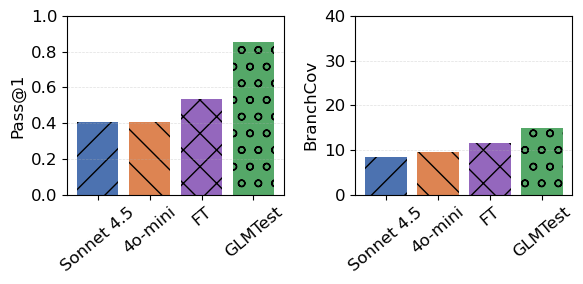}
    \caption{\texttt{Pass@1} and \texttt{BranchCov} when using branch-
targeted inference.} \vspace{-10pt}
    \label{fig:pass-and-cov}
\end{figure}

We evaluate the test suite's quality and coverage under the \texttt{GLMTest} inference procedure. We use \texttt{coverage.py} to enumerate feasible execution branches. Then, for each mechanism, we generate one test per branch and aggregate the resulting test cases into a single test suite. We consider the number of processed branches at $\delta = 1,000$, which is sufficient to cover \emph{all} branches for 41 out of 42 modules in the test set, providing ample branch-level coverage while keeping generation cost manageable. Then, we do the comparision under this shared pipeline in terms of \texttt{Pass@1} and \texttt{BranchCov}.

In this branch-targeted setting (Figure~\ref{fig:pass-and-cov}), \texttt{GLMTest} attains substantially higher \texttt{Pass@1} than the prompt-engineering baselines built on Claude-Sonnet-4.5 and GPT-4o-mini ($\sim 0.85$ vs. $0.41$), indicating that \texttt{GLMTest}'s test suites are substantially higher quality and more reliably executable. Also, \texttt{GLMTest} achieves the highest \texttt{BranchCov}, which strengthens its superior branch accuracy, indicating that a higher number of the targeted execution branches are actually exercised.

In addition, we consider a complementary setting in which \texttt{GLMTest} follows its inference pipeline while Claude-Sonnet-4.5 and GPT-4o-mini are prompted with the original TestGenEval prompt template, which does \emph{not} impose explicit branch targets and allows them to generate and explore freely. We include this setting for a fair comparison with the TestGenEval benchmark \cite{jaintestgeneval}, evaluating the functionality of generated test cases. In this setting, \texttt{GLMTest} still achieves the highest Pass@1 (i.e., 0.85 vs.\ 0.71 and 0.67 for Claude-Sonnet-4.5 and GPT-4o-mini, correspondingly), indicating that \texttt{GLMTest} consistently produces more reliable and executable test suites.



\subsection{RQ3: Ablation studies}

\begin{table}[t]
    \centering
    \small
    \begin{tabular}{llr}
        \toprule
        \textbf{Factor} & \textbf{Model Variant} & \textbf{\texttt{BranchAcc} $\uparrow$} \\
        \midrule
        \multirow{3}{*}{GNN structure}
            & GAT   & \textbf{0.502} \\
            & SAGE  & 0.465 \\
            & None (FT)  & 0.442 \\
        \midrule
        \multirow{2}{*}{Branch emb}
            & Node emb  & \textbf{0.502} \\
            & Graph emb & 0.399 \\
        \bottomrule
    \end{tabular}
    \caption{Ablation on GNN structure and branch embedding mechanism \texttt{GLMTest}.}
    \label{tab:gnn-ablation}
\end{table}

We conduct extensive ablation experiments to shed light on understanding the effects of each component of \texttt{GLMTest} on its performance. 

\textbf{GNN Structures.} We vary the GNN structure while keeping the backbone LLM and training data fixed, comparing our default GAT encoder against a GraphSAGE-based alternative. This experiment evaluates the effect of the message-passing architecture on the quality and contribution of the branch embeddings. Comparing \texttt{GLMTest} with its GraphSAGE-based alternative (Table~\ref{tab:gnn-ablation}) indicates that replacing the default GAT encoder with GraphSAGE leads to a drop in branch accuracy from $0.502$ to $0.465$, suggesting that the multi-head attention mechanism of GAT is better suited to capturing the heterogeneity in the code graphs. 


\textbf{Aggregation Mechanisms.} We examine how the branch embedding is constructed by changing the aggregation mechanism, as follows: \emph{(1) node-emb:} the embeddings of the nodes related to the targeted branch are concatenated; and \emph{(2) graph-emb}: node embeddings are first pooled (mean operator) into a single branch vector before being injected into the LLM. As in Table \ref{tab:gnn-ablation}, the node-level masking variant, which exposes all masked node embeddings directly to the LLM, achieves $0.502$ branch accuracy, whereas pooling these nodes into a single branch vector reduces performance to $0.399$. This indicates that preserving fine-grained structural information at the node level is vital for precise branch targeting, and motivates our choice of GAT with node-level masking as the default configuration for \texttt{GLMTest}.

\textbf{Model Size.} To assess the impact of model scale, we evaluate a larger backbone, Qwen2.5-Coder-14B-Instruct, under the same \texttt{GLMTest} training and inference pipeline. The 14B model achieves 49.3\% \texttt{BranchAcc} and 17.63\% \texttt{BranchCov}, compared to 50.2\% and 16.91\% for our default 7B model. These results suggest that explicit structural conditioning already provides strong branch-targeted reasoning at a moderate scale, while parameter scaling primarily benefits coverage breadth by enabling more diverse exploration of execution behaviors.

\textbf{Training Cost.}
We train \texttt{GLMTest} for $2{,}048$ optimization steps with a per-device batch size of $8$ and gradient accumulation of $32$, using a maximum sequence length of $8{,}192$. To optimize training speed, we leverage LoRA with rank $8$ and DeepSpeed on 4$\times$ A100 GB GPUs. Each GPU occupies only 50GB of memory, runs for roughly 48 wall-clock hours (about 192 GPU-hours in total), and processes approximately $5.37\times10^8$ tokens. This fine-tuning budget, i.e., \$119.56 on \texttt{Vast.ai}\footnote{\url{https://vast.ai/pricing}}, is modest, making it easy to adopt \texttt{GLMTest} in practices.

\section{Discussion}
\label{sec:discussion}


\textbf{Practical Use Cases.} \texttt{GLMTest} is well suited for practical use cases such as security analysis, where it can target high-risk code paths (e.g., input validation flagged by static analyzers) by generating concrete test suites that exercise these regions. More generally, \texttt{GLMTest} integrates naturally into fuzzing pipelines, supplying high-quality seeds to coverage-guided or mutation-based fuzzers and improving analysis depth and efficiency.


\textbf{Working with Other Languages.} Although our experiments focus on Python, adapting \texttt{GLMTest} to other languages is straightforward. Joern already supports code property graph extraction for multiple languages (e.g., C/C++), enabling the GNN component to operate without architectural changes, and modern code LLMs are multilingual. The primary challenge lies in data curation, which requires executing test suites in language-specific environments and recording the exercised branches.

\textbf{Mitigating the Limitation of CPG.} While \texttt{GLMTest} relies on static CPG extraction, syntactically present branches may be unreachable due to dead code or unsatisfiable runtime constraints. As a mitigation, dynamic analysis tools such as PyAnalyzer~\cite{jin2024pyanalyzer} can resolve library dependencies on demand and validate branch feasibility, and combining them with retrieval-based context expansion could further improve runtime condition resolution, which is a promising direction for future work.




\section{Conclusion}

We presented \texttt{GLMTest}, a novel graph-enhanced language modeling framework that treats feasible execution branches as explicit test-generation targets. By integrating structural and textual information, \texttt{GLMTest} enables structure-aware test case generation. Our experimental results show that \texttt{GLMTest} built on the Qwen2.5-Coder-7B-Instruct model achieves high branch accuracy and executability, while achieving competitive branch coverage compared with state-of-the-art commercialized LLMs (Claude-Sonnet-4.5 and GPT-4o-mini), highlighting the advantages of \texttt{GLMTest}. 


\section*{Limitations}

While \texttt{GLMTest} improves branch-targeted test case generation on our benchmark, it has several limitations. First, our current model is trained on a relatively small set of projects from TestGenEval and does not yet demonstrate strong cross-project generalization. Extending training to a broader and more diverse corpus of repositories is a natural next step. Second, the branch-targeted inference pipeline can become expensive on very large, highly modular systems with thousands of feasible branches. In such settings, applying \texttt{GLMTest} to every branch is impractical. This is a problem for all testing approaches - not just \texttt{GLMTest}- and the method is better viewed as a targeted tool for a subset of critical branches. This limitation also suggests future work on principled branch prioritization, for example, by combining \texttt{GLMTest} with a static security risk detection mechanism~\cite{lekssays2025llmxcpg, li2025vulpo}.

\section*{Acknowledgments}

This research was supported by the National Science Foundation (NSF) under Grant No. CNS 2237328 and DGE
2043104, and the Grace Hopper AI Research Institute.

\bibliography{main}

\appendix

\section{Experimental details}

\subsection{Dataset processing details}
\label{appx:dataset}
From each project in TestGenEval, we first decompose the available test suites into individual test cases. For every test case, we statically remove unused imports to simplify the context and reduce opportunities for the model to hallucinate spurious dependencies. We then execute each test case inside its official Docker environment and collect \emph{branch} information using \texttt{coverage.py}\footnote{\url{https://coverage.readthedocs.io/en/7.13.0/}} in branch-coverage mode. Test cases that fail due to environment issues, timeouts exceeding 60 seconds per Jain et al. \cite{jaintestgeneval} or nondeterministic behavior, are discarded, and we run each remaining test once to obtain a stable branch set. Each qualified test case and its associated set of executed branches form a data point, yielding (program, branch, test case) triples as described in Section~\ref{sec:train}. Projects that yield fewer than 15 valid triples after filtering are removed (4 of the 11 repositories are discarded), as they provide little signal and complicate stratified sampling. After preprocessing, we obtain 7 projects, 45,831 unique triples (program, branch, test case) (see Table~\ref{tab:dataset-stats}). For evaluation, we reserve 1{,}489 test instances, sampled uniformly across the remaining projects to avoid project skew, and use the rest for training and validation; within this split, projects are shared across splits, but individual test cases are disjoint, so our results primarily measure generalization to unseen test cases within the same set of projects.

\subsection{Implementation}
\label{appx:implementation}

\textbf{Additional details of node's features.} Each CPG node is annotated with source-location metadata (file path, start line, end line) and a set of categorical attributes (e.g., syntactic type, role in the AST or control/data flow). For node text features, we encode the textual content associated with each node (code snippet and identifier context, excluding comments and docstrings) using the \texttt{Salesforce/codet5p-110m-embedding} pretrained model. We use the CodeT5p encoder to obtain a 768-dimensional embedding. We then concatenate this code embedding with a 4-dimensional label-encoded vector of categorical node attributes to obtain a 772-dimensional per-node feature vector, thereby combining rich pre-trained code semantics with lightweight structural metadata.

\textbf{Additional details of branch mask construction.} To construct branch masks, for each executed branch, we obtain the corresponding set of executed line numbers and align them with CPG nodes via their source-location intervals: a node is marked as relevant (mask value 1) if its line range intersects the executed line set, and irrelevant (mask value 0) otherwise. In rare cases where no CPG node aligns with an executed branch, we set the structural embedding and the prompt input to \textit{"Not available"}. This line-level alignment provides a direct, interpretable mapping from dynamic execution to static structure, enabling \texttt{GLMTest} to focus on subgraphs along the targeted execution path while remaining compatible with standard coverage tooling.

\subsection{Baseline settings}
\label{appx:baseline-details}

We compare \texttt{GLMTest} against prior work on automated test case generation and two dataset-compatible LLM baselines. Recent systems include ASTER~\cite{pan2025aster}, CodaMOSA~\cite{10172800}, ACH~\cite{harman2025mutation}, SymPrompt~\cite{ryan2024code}. Among these, only ASTER and CodaMOSA directly target Python unit tests, but both are implemented as Pynguin-based pipelines that assume locally importable modules and direct filesystem access to the project under test. In contrast, TestGenEval executes each repository inside an isolated Docker container with its own entrypoint and dynamically configured \texttt{PYTHONPATH}, and does not expose the Pynguin-style project orchestration interface. In our attempts to run ASTER and CodaMOSA on TestGenEval, we were unable to make their Pynguin-based harness discover and import the correct modules inside the official containers without substantial re-engineering of their toolchains, which we consider out of scope for this work.\footnote{We therefore report no ASTER/CodaMOSA numbers on TestGenEval; our code release will document the incompatibility and configuration attempts.}

Within these constraints, we use two reproducible LLM baselines that share the same decoding budget and evaluation protocol as \texttt{GLMTest}. (i) \emph{Prompt-only LLM (PE).} Following the TestGenEval setting, we query LLMs with a fixed prompt template provided by TestGenEval that includes the program source and a textual description of the target branch (its line range and correct order of lines executed as in Figure \ref{fig:branch-example}), but no CPG-derived features. For each instance, we generate a single test case ($k = 1$) using temperature 0.2 and greedy decoding strategy, so that Pass@1 is directly comparable across models. (ii) \emph{Text-only fine-tuning (FT).} We fine-tune the same backbone LLM as \texttt{GLMTest} on our (program, branch, test case) triples, but remove the GNN and represent the branch set purely as text (a serialized list of executed line ranges) concatenated with the program source and instruction prompt, capped at 8192 tokens. FT therefore has access to branch information only through this textual description, without any explicit graph structure or relational context, providing a strong non-structural baseline that isolates the contribution of CPG-based conditioning in \texttt{GLMTest}.

\section{Related work}
\label{appx:related-work}

\paragraph{LLMs for test case generation.}
Recently, LLMs have been applied to software testing to produce readable, executable test suites and improve coverage~\cite{tufano2021unittestcasegeneration}. Existing approaches generally fall into two categories: fine-tuning and prompt engineering. Fine-tuning methods train on curated code–test pairs to specialize LLMs for test case generation~\cite{tufano2021unittestcasegeneration, ALAGARSAMY2024107565, 10.1109/ASE56229.2023.00193}, whereas prompt-based methods keep the LLM frozen and construct structured prompts from extracted program features (e.g., signatures, control-flow summaries) to guide coverage-oriented generation~\cite{10329992, 10.1145/3661167.3661216, 10.1145/3663529.3663801, DAKHEL2024107468}. These techniques have shown promising gains in global coverage, but they do not explicitly represent or optimize for specific execution branches, which limits their effectiveness in scenarios where developers or security analysts need to exercise particular high-risk paths.


\paragraph{Combining CPGs with LLMs.}
Recent works~\cite{lekssays2025llmxcpg, chen2025bridging, liu2025codexgraph} have begun to combine CPGs with LLMs for downstream code understanding and analysis tasks, typically treating the graph as a knowledge source to enrich the prompt or as a generic encoder whose outputs are consumed at the sequence level. Lekssays et al. \cite{lekssays2025llmxcpg} leverage the CPG to generate a code slice from the codebase, keeping relevant code lines, and prompt the LLMs with the code slice for vulnerability detection. Chen et al. \cite{chen2025bridging} proposed a framework that incorporates CPG-derived node features into the LLM's forward pass to enhance code understanding. However, existing works usually extract code snippets guided by the graph without explicitly encoding the underlying structural relationships into branch-specific representations. 


\section{Use of AI Assistants}

In this work, we leverage the help of AI assistants to facilitate the work as follows. For the literature search, we use the Google Scholar Labs agent to find relevant works. However, all citations are manually checked and selected by the authors. To implement the project, we use Copilot, equipped with Claude-Sonnet-4.5, as a coding assistant to edit the code. Nevertheless, all experimental designs, algorithmic choices, and executions are conducted manually by the authors. For writing, we used GPT-5.2 as an assistant purely with the language of the paper. The problem formulation, technical contributions, and empirical analysis were conducted by the authors.

\section{Supplemental Results}
\label{appx:supplemental_results}

\begin{table}[!htp]
    \centering
    \small
    \begin{tabular}{lrrr}
        \toprule
        \textbf{Project} & \textbf{\# Train} & \textbf{\# Test} & \textbf{\# Stars} \\
        \midrule
        \texttt{astropy}       & 2{,}714 & 6   & 5.0k \\
        \texttt{django}        & 19{,}364 & 690 & 86.3k \\
        \texttt{xarray}        & 1{,}615 & 23  & 4.0k \\
        \texttt{pytest}        & 2{,}617 & 93  & 13.4k \\
        \texttt{scikit-learn}  & 5{,}918 & 319 & 64.4k \\
        \texttt{sympy}         & 8{,}640 & 213 & 14.2k \\
        \midrule
        \textbf{Total}         & 40{,}868 & 1{,}344 & -- \\
        \bottomrule
    \end{tabular}
    \caption{Number of training and test data points per repository after preprocessing, along with approximate GitHub star counts (as of late 2025).}
    \label{tab:dataset-stats}
\end{table}

\begin{figure}[t]
    \centering
    \begin{minipage}{\linewidth}
    \small
    \begin{lstlisting}[basicstyle=\ttfamily\footnotesize, frame=single]
# INSTRUCTION: You are an AI agent that
generates executable Python test cases
targeting a specific execution branch 
of a module.

Inputs:
- Module source: source code of the
target module (Could be truncated to 
related line only).
- Execution branch information: the 
lines of the target module executed.
- Module path: a valid, importable path 
from the PYTHONPATH directory.
- Code Property Graph (CPG) embeddings
(Optional): semantic and structural 
information about the code elements 
related to the branch.

Tasks:
1. Generate a runnable Python test file 
that executes the specified branch of 
the module.
2. Include meaningful assertions that 
confirm correct behavior and should pass 
for the given branch.
3. Output only the final, runnable 
Python test code, no explanations 
or reasoning text.

Requirements:
- All imports must be valid and 
correspond to existing modules; do not 
invent or hallucinate any packages.
- Use standard testing practices (unit-
test, pytest, or assert statements).
- Keep the code clear, minimal, 
and maintainable.
---------------------------------------

# INPUTS:

## Module Source: <Input>

## Execution Branches Information 
(Line to Line executed): <Input>

## Module Path: <Input>

## Code Property Graph 
(CPG) Node Embeddings: <Input>

## Here's how to import the target 
module: <Input>
\end{lstlisting}
    \end{minipage}
    \caption{Prompt template used by \texttt{GLMTest} to instruct the LLM to generate branch-targeted Python test cases.}
    \label{fig:prompt-template}
\end{figure}

\end{document}